\documentclass[preprint,NumberedRefs]{JASAnew}
\usepackage{tikz}
\usepackage[capitalise]{cleveref}

\newcommand{\mat}{\left[\begin{matrix}}
\newcommand{\emat}{\end{matrix}\right]}
\newcommand{\eps}{{\varepsilon}}
\newcommand{\avg}[1]{\left\langle {#1} \right\rangle}

\newcommand{\tens}{\mathsf}

\begin{document}

\title[One-dimensional homogenization]{Homogenization of one-dimensional layered and graded structures}
\author{Michael B.\ Muhlestein}
\email{Michael.B.Muhlestein@usace.army.mil}
\affiliation{U.\ S.\ Army Engineer Research and Development Center, 72 Lyme Rd., Hanover, NH 03755}
\author{Alexei T. Skvortsov}
\affiliation{Defence Science and Technology Group, 506 Lorimer Street, Fishermans Bend, VIC 3207, Australia}

\date{\today}

\begin{abstract}
	The homogenization of one-dimensional acoustic or elastic structures of finite extent is considered.  A new homogenization method based on transfer matrices is derived.  The new homogenization method may account for variable cross sectional area and for Willis coupling, which couples the stress-strain and momentum-velocity constitutive relations.  The homogenization method is then demonstrated by considering acoustic waves normally incident upon a rigidly-backed double-layered wall and plane waves propagating in a duct with a section of exponentially-growing cross-sectional area.
\end{abstract}

\maketitle

\section{Introduction}

The study of acoustic metamaterials hinges on the ability to determine the effective material properties of a system, also known as homogenization.  Homogenization of one-dimensional systems has been extensively studied in the static case.  Analytical methods of homogenization can be especially useful as efficient design tools due to the fact that they provide explicit results.  Examples of previous analytical homogenization methods in one dimension include averaging equations for quasi-static deformations, analyzing periodic systems of layered media,\cite{white1955,rytov1956,brekhovskikh2012,smith2011,liu2009,willis2009,brekhovskikh2012} and collective modes in the systems of lumped elements.\cite{bobrovnitskii2014,jimenez2016}  As an example important to the present work, Kutsenko, \emph{et al}.\ used a $4\times4$ transfer matrix to describe propagation in an infinite, one-dimensional, periodic, layered piezoelectric medium.\cite{kutsenko2015}  They were able to homogenize the system both in the quasi-static limit and for finite frequencies by analyzing the dispersion relation of the propagated waves.  While their approach is quite general, except for in the quasi-static limit it assumes an infinitely-periodic system and cannot account for finite sizes of materials.  Finite-sized systems materials can be important for designing inclusions for multiscale homogenization methods in periodic\cite{sieck2015} and non-periodic media\cite{baird1999,muhlestein2016} and for analyzing the behavior of composite plates.\cite{zhu2015}  In addition, it assumes that all layers are of infinite lateral extent and therefore cannot account for one-dimensional ducts with variable cross-sectional area.  The purpose of this paper is to present a related but alternative homogenization method to that of Kutsenko, \emph{et al}.\ that accurately homogenizes one-dimensional systems that may include finite sizes and variable cross-sectional areas, though not piezoelectric properties.

The outline of the paper is as follows.  In Sec.~\ref{transferMatrix} the alternative transfer matrix homogenization method is presented in both discrete and continuous representations.  Section~\ref{example} provides examples of the homogenization method.  Finally, Sec.~\ref{conclusions} summarizes the conclusions.

\section{Transfer Matrix Homogenization}\label{transferMatrix}

The Willis constitutive equations in one dimension may be written as\cite{muhlestein2017a}
\begin{align}
	-p &= \kappa\eps + \psi^{(1)} \dot v, &
	\mu &= \rho v + \psi^{(2)}\dot\eps,
\end{align}
where $p$ is the acoustic pressure, $\eps$ is the volume strain, $\mu$ is the momentum density, $v$ is the particle velocity, and over-dots denote time derivatives.  The material properties are the bulk modulus $\kappa$, the mass density is $\rho$, and the Willis coupling is represented by $\psi^{(1)}$ and $\psi^{(2)}$.  For passive and causal systems the Willis coupling coefficients are equal,\cite{muhlestein2016a} i.e., $\psi^{(1)}=\psi^{(2)}$.  The constitutive equations supplement the dynamic equation and the definition of the strain rate:
\begin{align}
	\dot\mu &= -p', &
	\dot\eps &= v',
\end{align}
where the primes denote spatial derivatives.  Combining these equations together leads to the standard wave equation with the wave speed $c=\sqrt{\kappa/\rho}$.\cite{willis2009}  Assuming time-harmonic motion ($e^{-i\omega t}$ time convention) leads to the conclusion that the wavenumber $k=\omega/c$.

While the analysis presented here and below assumes all materials are fluids, it is worthwhile to note that in isotropic solids the longitudinal and shear waves are independent of each other, and in one dimension there is no mathematical distinction between these elastic waves and fluid waves.  Thus if $G$ is the shear modulus, replacing the bulk modulus $\kappa$ with the plane wave modulus $\kappa+4G/3$ yields the same results for longitudinal elastic waves and replacing $\kappa$ with $G$ yields the same results for shear elastic waves.  Note that this correspondence is only valid for one-dimensional propagation, as the interface conditions become coupled for oblique incidence.

Given an inhomogeneous domain $\Omega=(a,b)$ where $k(b-a)\equiv kL\ll 1$, these constitutive equations may be used to define the effective material properties of the domain.  These effective material properties may be written as
\begin{subequations}\label{effDef}
\begin{align}
	\kappa_\text{eff} &\equiv \left.i\omega\frac{\avg{p}}{\avg{v'}}\right|_{\avg{v}=0}, & \rho_\text{eff} &\equiv \left.\frac{1}{i\omega}\frac{\avg{p'}}{\avg{v}}\right|_{\avg{v'}=0}, \\ \psi_\text{eff}^{(1)} &\equiv \left. \frac{1}{i\omega}\frac{\avg{p}}{\avg{v}} \right|_{\avg{v'}=0}, & \psi_\text{eff}^{(2)} &\equiv \left. \frac{1}{i\omega}\frac{\avg{p'}}{\avg{v'}} \right|_{\avg{v}=0}.
\end{align}
\end{subequations}
These averages may be written in terms of the field quantities at the edges of the domain.  If the domain is $\Omega=(a,b)$, where $b-a=L>0$, then the average fields may be written as
\begin{subequations}\label{avgDef}\begin{align}
	\avg{p} &\approx \frac{p(a)+p(b)}{2}, &
	\avg{v} &\approx \frac{v(a)+v(b)}{2}, \\
	\avg{p'} &\approx \frac{p(b)-p(a)}{L}, &
	\avg{v'} &\approx \frac{v(b)-v(a)}{L}.
\end{align}\end{subequations}
The fields at the edges of the domain $\Omega$ are generally related by an ABCD transmission matrix
\begin{equation}\label{ABCDdef}
	\mat p(a) \\ v(a) \emat = \mat ~~A~~ & ~~B~~ \\ C & D \emat \mat p(b) \\ v(b) \emat.
\end{equation}
Using Eqs.~\eqref{effDef}--\eqref{ABCDdef} the effective material properties may then be written as
\begin{subequations}
\begin{align}
	\kappa_\text{eff} &= -i\omega L\frac{A+D+1+(AD-BC)}{4C}, \\ \rho_\text{eff} &= \frac{1}{-i\omega L}\frac{A+D-1-(AD-BC)}{C}, \\
	\psi_\text{eff}^{(1)} &= \frac{1}{-i\omega}\frac{D-A-1+(AD-BC)}{2C}, \\ \psi_\text{eff}^{(2)} &= \frac{1}{-i\omega}\frac{D-A+1-(AD-BC)}{2C}.
\end{align}
\end{subequations}
Thus the effective material properties of a one-dimensional system may be obtained with a knowledge of the systems $ABCD$ tranmission matrix.

If the structure is passive and reciprocal, then the determinant of the $ABCD$ matrix is $AD-BC=1$.  In this case the effective material properties simplify to the expressions
\begin{subequations}\label{effPropsPR}
\begin{align}
	\kappa_\text{eff} &= -i\omega L\frac{A+D+2}{4C}, \\ \rho_\text{eff} &= \frac{1}{-i\omega L}\frac{A+D-2}{C}, \\
	\psi_\text{eff}^{(1)} &= \frac{1}{-i\omega}\frac{D-A}{2C} = \psi_\text{eff}^{(2)}\equiv\psi_\text{eff}.
\end{align}
\end{subequations}

\begin{figure}
\centering
\includegraphics{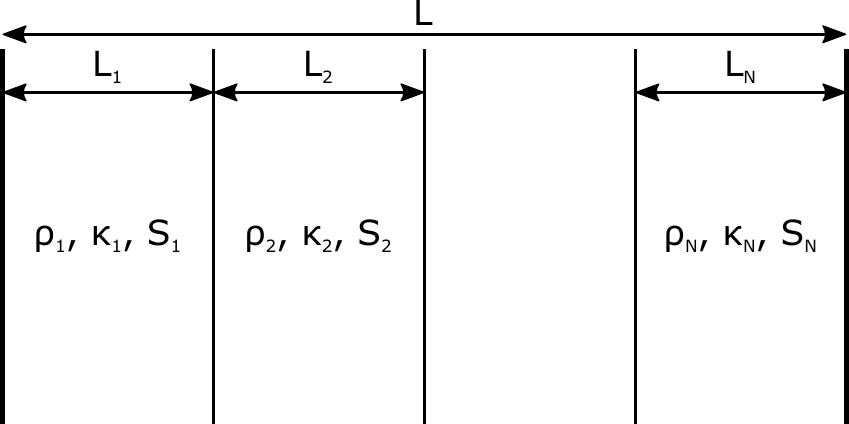}
	\caption{\label{layeredMedium}Schematic of a finite one-dimensional layered acoustical medium of length $L$.  Each layer has an associated mass density $\rho$, bulk modulus $\kappa$, and cross-sectional area $S$.}
\end{figure}

Consider a one-dimensional layered material of length $L$ as shown in Fig.~\ref{layeredMedium}.  The the system consists of $N$ layers and the $n^\text{th}$ layer has length $L_n$, has mass density $\rho_n$, bulk modulus $\kappa_n$, and (for ducts) cross-sectional area $S_n$.  The acoustic pressure and the volume velocity on the left-hand side of the $n^\text{th}$ layer, $p_{n-1}$ and $q_{n-1}$ respectively, may be related to the acoustic pressure and volume velocity on the right-hand side, $p_n$ and $v_n$, by a standard $ABCD$ matrix:
\begin{equation}
	\mat p_{n-1} \\ q_{n-1} \emat = \mat \cos(k_nL_n) & -iZ_n\sin(k_nL_n) \\ -\frac{i}{Z_n}\sin(k_nL_n) & \cos(k_nL_n) \emat \mat p_n \\ q_n \emat,
\end{equation}
where $k_n=\omega\sqrt{\rho_n/\kappa_n}$ and $Z_n=\sqrt{\rho_n\kappa_n}/S_n$ are the wavenumber and acoustic impedance of the $n^\text{th}$ layer.  These expressions may be combined to relate the fields at the left-hand side of the entire structure to the fields at the right-hand side as
\begin{equation}
	\mat p_0 \\ q_0 \emat = \left(\prod_{n=1}^N\mat \cos(k_nL_n) & -iZ_n\sin(k_nL_n) \\ -\frac{i}{Z_n}\sin(k_nL_n) & \cos(k_nL_n) \emat \right) \mat p_N \\ a_N \emat \equiv \mat ~~A'~~ & ~~B'~~ \\ C' & D' \emat \mat p_N \\ q_N \emat.
\end{equation}
While this analysis does indeed yield an $ABCD$ matrix, it is written for the volume velocity rather than the particle velocity.  Defining $S_\text{ref}$ as a reference or effective cross-sectional area it is straightforward to find that the elements of the $ABCD$ matrix in terms of the particle velocity may be written as $A = A'$, $B = B'S_\text{ref}$, $C = C'/S_\text{ref}$, and $D = D'$.

For $\omega$ small enough such that $k_nL_n\ll\pi/2$ the $N$ $2\times2$ matrices
\begin{equation}
	\tens A_n \equiv \mat \cos(k_nL_n) & -iZ_n\sin(k_nL_n) \\ -\frac{i}{Z_n}\sin(k_nL_n) & \cos(k_nL_n) \emat
\end{equation}
may be expanded in a matrix series as
\begin{equation}\label{approxABCD}
	\tens A_n = \tens I -ik_nL_n\tens D_n - \frac{(k_nL_n)^2}{2}\tens I + O([k_nL_n]^3),
\end{equation}
where $\tens I$ is the identity $2\times2$ matrix and
\begin{equation}
	\tens D_n \equiv \mat ~~0~~ & Z_n \\ \frac{1}{Z_n} & ~~0~~ \emat.
\end{equation}
Define $\eps$ as the largest value of $k_nL_n\equiv\theta_n$, such that $k_nL_n$ is of order $\eps$ for all $n$.  Then, using the results from \cref{appendixA} we find that
\begin{equation}
	\prod_{n=1}^N\tens A_n \approx \tens I - i\sum_{n=1}^N k_nL_n\tens D_n - \frac{1}{2}\sum_{n=1}^N (k_nL_n)^2\tens I - \sum_{n=1}^{N-1}\sum_{m=n+1}^N k_nL_nk_mL_m\tens D_n\tens D_m.
\end{equation}
The elements of the composite $ABCD$ may then be approximated as
\begin{subequations}\label{totABCDapprox}\begin{align}
	A' &\approx 1 - \frac{1}{2}\sum_{n=1}^N (k_nL_n)^2 - \sum_{n=1}^{N-1}\sum_{m=n+1}^N k_nL_nk_mL_m\frac{Z_n}{Z_m}, \\
	B' &\approx {\color{white}0} -i \sum_{n=1}^N k_nL_nZ_n, \\
	C' &\approx {\color{white}0} -i \sum_{n=1}^N \frac{k_nL_n}{Z_n}, \\
	D' &\approx 1 - \frac{1}{2}\sum_{n=1}^N (k_nL_n)^2 - \sum_{n=1}^{N-1}\sum_{m=n+1}^N k_nL_nk_mL_m\frac{Z_m}{Z_n}.
\end{align}\end{subequations}
Since $k_nZ_n=\omega\rho_n/S_n$ and $k_n/Z_n=\omega S_n/\kappa_n$, the effective material properties may then be written to lowest order as
\begin{subequations}\label{effProps}\begin{align}
	\frac{1}{\kappa_\text{eff}} &= \frac{1}{S_\text{ref}}\avg{ \frac{S_n}{\kappa_n}}_{\!\!n}, \label{effK} \\
	\rho_\text{eff} &= S_\text{ref}\avg{\frac{\rho_n}{S_n}}_{\!\!n}, \label{effR} \\
	\psi_\text{eff} &= \frac{\kappa_\text{eff}}{2}\avg{\sum_{m=n+1}^N L_m\left(\frac{\rho_mS_n}{\kappa_nS_m}-\frac{\rho_nS_m}{\kappa_mS_n}\right)}_{\!\!n}, \label{effP}
\end{align}\end{subequations}
where
\begin{equation}
	\avg{~\cdot~}_{\!n} \equiv \frac{1}{L}\sum_{n=1}^NL_n[~\cdot~]
\end{equation}
is the spatial average operator.

There are multiple interesting features of the predicted effective material properties in Eqs.~\eqref{effProps}.  First every term depends on the stiffness, meaning that simple averages of the mass density and Willis coupling coefficient are inaccurate.  Another point of interest is that the stiffness always appears in summations as its inverse, the compressibility.  Thus, layers with very low stiffness tend to dominate the overall response of the system.  The Willis coupling coefficient approaches a real constant, even in the zero-frequency limit.  Since the summand of the Willis coupling coefficient is odd with respect to $m$ and $n$, symmetric systems will not display any Willis coupling.  Additionally, two layers, $m$ and $n$, do not contribute to the Willis coupling if $\rho_mS_n/\kappa_nS_m = \rho_nS_m/\kappa_mS_n$, which reduces to equality of the acoustic impedances squared, $(Z_m/S_m)^2=(Z_n/S_n)^2$.

A one-dimensional system with continuously varying properties may be treated with the above framework by letting $P_n\rightarrow P(x)$, where $P\in\{\rho,\kappa,S\}$, and $L_n\rightarrow \mathrm dx$.  In this case the effective material properties become
\begin{subequations}\label{effCont}\begin{align}
	\frac{1}{\kappa_\text{eff}} &= \frac{1}{S_\text{ref}}\int_{0}^L \mathrm dx \frac{S(x)}{\kappa(x)} \equiv \frac{1}{S_\text{ref}}\avg{\frac{S(x)}{\kappa(x)}}, \label{effKcont} \\
	\rho_\text{eff} &= \frac{S_\text{ref}}{L}\int_{0}^L \mathrm dx\frac{\rho(x)}{S(x)} \equiv S_\text{ref}\avg{\frac{\rho(x)}{S(x)}}, \label{effRcont} \\
	\psi_\text{eff} &= \frac{\kappa_\text{eff}}{2}\avg{\int_x^L\mathrm dy\left(\frac{\rho(y)S(x)}{\kappa(x)S(y)}-\frac{\rho(x)S(y)}{\kappa(y)S(x)}\right)}. \label{effPcont}
\end{align}\end{subequations}

\section{Examples}\label{example}

\subsection{Finite System With Discrete Layers}

\begin{figure}
	\centering
	\includegraphics{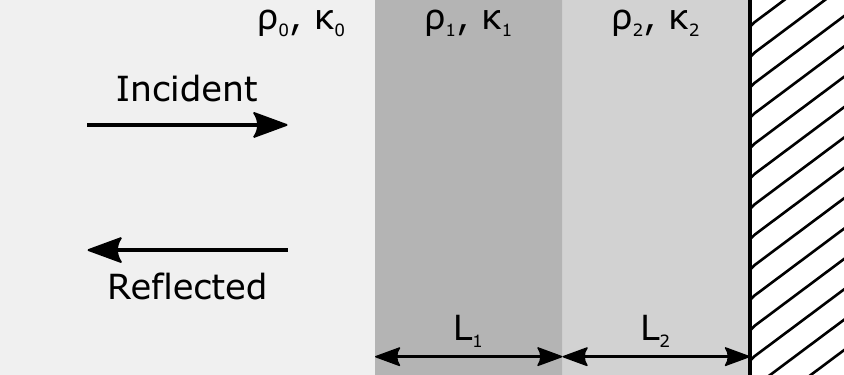}
	\caption{\label{discreteSchematic}Schematic of a plane wave normally incident upon a bi-layer wall with a rigid backing.}
\end{figure}

Consider the reflection problem described schematically in \cref{discreteSchematic}.  The background material has mass density and bulk modulus of $\rho_0$ and $\kappa_0$, and the $j^\text{th}$ layer has the properties $\rho_j$ and $\kappa_j$ and is of width $L_j$.  The acoustic pressure field in the background medium for a normally incident plane wave may then be written as
\begin{equation}
	p_0 = A_0\left[e^{ik_0z}+Re^{-ik_0z}\right],
\end{equation}
where $A_0$ is the amplitude of the incoming wave, $R$ is the reflection coefficient, and $k_0=\omega\sqrt{\rho_0/\kappa_0}$ is the incident wavenumber.  It is then straightforward to apply continuity of particle velocity and acoustic pressure at the interfaces and show that the reflection coefficient may be written $R = (1-\zeta)/(1+\zeta)$ where $\zeta$ is a normalized input impedance given by
\begin{equation}
	\zeta = -i\frac{Z_0}{Z_1}\frac{\tan(k_1L_1)+\tan(k_2L_2)\frac{Z_1}{Z_2}}{1-\tan(k_1L_1)\tan(k_2L_2)\frac{Z_1}{Z_2}},
\end{equation}
$Z_i=\sqrt{\rho_i\kappa_i}$, and $k_i=\omega\sqrt{\rho_i/\kappa_i}$.  For low frequencies the normalized input impedance may be approximated as
\begin{equation}
	\zeta \approx -i\omega Z_0\left[\left( \frac{L_1}{\kappa_1}+\frac{L_2}{\kappa_2} \right) + \frac{\omega^2}{3}\left(\frac{\rho_1}{\kappa_1^2}L_1^3 + 3\frac{\rho_1}{\kappa_1\kappa_2}L_1^2L_2 + 3\frac{\rho_1}{\kappa_2^2}L_1L_2^2 + \frac{\rho_2}{\kappa_2^2}L_2^3\right)\right].
\end{equation}

The bi-layer wall may be approximated at low frequencies by a single layer of width $L=L_1+L_2$ with effective material properties as prescribed by Eqs.~\eqref{effProps}.  Since all cross sectional areas are equal we thus obtain
\begin{subequations}
\begin{align}
	\kappa_\text{eff} &= \frac{L_1+L_2}{\frac{L_1}{\kappa_1}+\frac{L_2}{\kappa_2}}, \\
	\rho_\text{eff} &= \frac{L_1\rho_1 + L_2\rho_2}{L_1+L_2}, \\
	\psi_\text{eff} &= \frac{L_1L_2}{2}\frac{\frac{\rho_2}{\kappa_1} - \frac{\rho_1}{\kappa_2}}{\frac{L_1}{\kappa_1}+\frac{L_2}{\kappa_2}}.
\end{align}
\end{subequations}
In line with the above comments $\psi_\text{eff}=0$ if the layers have equal impedance.  As noted above the acoustic pressure and particle velocity in the Willis layer may be described by the wave equation with the standard wave speed.  Then, the acoustic fields may be written in terms of trigonometric functions as
\begin{subequations}
\begin{align}
	p_\text{eff} &= A_0\left[ A_1\cos(k_\text{eff}(L-z)) + B_1\sin(k_\text{eff}(L-z)) \right], \\
	v_\text{eff} &= \frac{A_0/iZ_\text{eff}}{1+W^2_\text{eff}}\left[ (W_\text{eff}A_1-B_1)\cos(k_\text{eff}(L-z)) + (A_1+W_\text{eff}B_1)\sin(k_\text{eff}(L-z)) \right],
\end{align}
\end{subequations}
where $Z_\text{eff}=\sqrt{\rho_\text{eff}\kappa_\text{eff}}$ is the effective characteristic impedance and $W_\text{eff}=\omega\psi_\text{eff}/Z_\text{eff}$ is the effective asymmetry factor (a non-dimensional measure of the importance of Willis coupling to total impedance\cite{muhlestein2017a}).  Requiring the backing to be rigid leads to the requirement $B_1=W_\text{eff}A_1$.  Then matching the pressure and particle velocity at $z=0$ leads to the equations
\begin{subequations}
\begin{align}
	1+R &= A_1\left[\cos(k_\text{eff}L) + W_\text{eff}\sin(k_\text{eff}L)\right], \\
	\frac{1}{Z_0}\left[ 1 - R \right] &= \frac{A_1}{iZ_\text{eff}}\sin(k_\text{eff}L),
\end{align}
\end{subequations}
which combine to yield $R = (1-\zeta_\text{eff})/(1+\zeta_\text{eff}),$ where the effective normalized input impedance is given by
\begin{equation}
	\zeta_\text{eff} = -i\frac{Z_0}{Z_\text{eff}}\frac{\tan(k_\text{eff}L)}{1 + W_\text{eff}\tan(k_\text{eff}L)}.
\end{equation}
For very small frequency we may then approximate
\begin{align}
	\frac{i\zeta_\text{eff}}{\omega Z_0} &\approx \frac{L}{\kappa_\text{eff}} + \frac{\omega^2}{3}\left( \frac{\rho_\text{eff}}{\kappa_\text{eff}^2}L^3 - 3\frac{\psi_\text{eff}}{\kappa_\text{eff}^2}L^2 \right) \notag \\
	=& \left(\frac{L_1}{\kappa_1}+\frac{L_2}{\kappa_2}\right) + \frac{\omega^2}{3}\left( \frac{\rho_1}{\kappa_1^2}L_1^3 + \left[ \frac{7}{2}\frac{\rho_1}{\kappa_1\kappa_2} - \frac{1}{2}\frac{\rho_2}{\kappa_1^2} \right]L_1^2L_2 + \left[ \frac{5}{2}\frac{\rho_1}{\kappa_2^2} + \frac{1}{2}\frac{\rho_2}{\kappa_1\kappa_2} \right]L_1L_2^2 + \frac{\rho_2}{\kappa_2^2}L_2^3 \right).
\end{align}
The difference between the low-frequency approximations of $\zeta$ and $\zeta_\text{eff}$ is
\begin{equation}
	\zeta - \zeta_\text{eff} = iZ_0\frac{\omega^3}{6}\frac{Z_1^2-Z_2^2}{\kappa_1\kappa_2}L_1L_2\left( \frac{L_1}{\kappa_1} - \frac{L_2}{\kappa_2} \right).
\end{equation}
This residual may be explicitly made zero in the case that $L_1/\kappa_1=L_2/\kappa_2$.  Thus, the effective material yields the same normalized input impedance as the full case to $O(\omega^3)$.  Note that if $\psi_\text{eff}$ were neglected then the difference would yield the error
\begin{equation}
	\zeta - \left.\zeta_\text{eff}\right|_{\psi_\text{eff}=0} = iZ_0\frac{\omega^3}{6}\frac{Z_1^2-Z_2^2}{\kappa_1\kappa_2}L_1L_2\left( -2\frac{L_1}{\kappa_1} - 4\frac{L_2}{\kappa_2} \right),
\end{equation}
which is still $O(\omega^3)$, but is greater error magnitude than the case where $\psi_\text{eff}$ is included.  Since $L_1/\kappa_1$ and $L_2/\kappa_2$ are both strictly positive, it becomes apparent that there is no way to reduce the $O(\omega^3)$ error to zero given $Z_1\ne Z_2$ without accounting for Willis coupling.  An analysis of the $O(\omega^5)$ error (not presented here) exhibits a similar behavior.  Thus, while neglecting Willis coupling in the effective layer provides an accurate reflection coefficient in the quasi-static limit, as the frequency increases Willis coupling becomes more important.

\subsection{Finite Duct With an Embedded Exponential Horn}

\begin{figure}
	\includegraphics{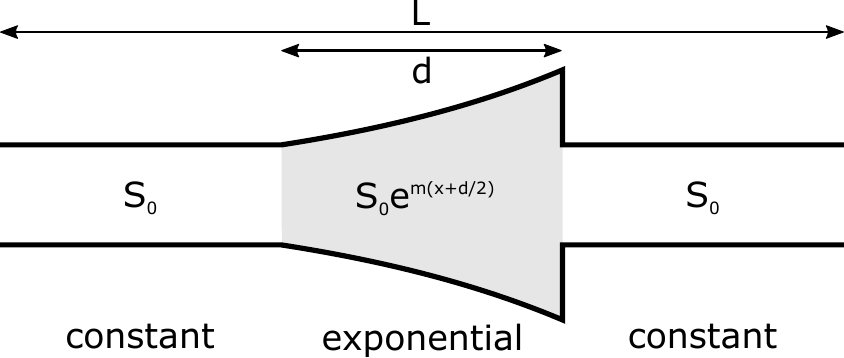}
	\caption{\label{hornSchematic}Schematic of a uniform circular duct with cross-sectional area $S_0$ of length $L$ with a small section of exponentially growing cross-sectional area of length $d$ embedded in the center.}
\end{figure}

Consider a uniform circular duct with cross-sectional area $S_0=S_\text{ref}$ of length $L$ with a small section of exponentially growing cross-sectional area of length $d$ centered in the duct, as shown in \cref{hornSchematic}.  In this case the cross-sectional area may be written as
\begin{equation}
	S(x) = S_0 \begin{cases} e^{m(x+d/2)} & -d/2<x<d/2 \\ 1 & \text{else} \end{cases}.
\end{equation}
The mass density and bulk modulus inside the duct are $\rho_0$ and $\kappa_0$.  Then, using Eqs.~\eqref{effCont}, we obtain
\begin{subequations}\label{fullTube}\begin{align}
	\kappa_\text{eff} &= \kappa_0\left[1-\phi+\frac{\phi}{md}\left(e^{md} - 1\right)\right]^{-1}, \\
	\rho_\text{eff} &= \rho_0\left[ 1-\phi+\frac{\phi}{md}\left(1 - e^{-md}\right) \right], \\
	\psi_\text{eff} &= \phi\frac{\rho_0}{m}\frac{1 - \frac{\sinh(md)}{md}}{1-\phi + \frac{\phi}{md}\left(e^{md}-1\right)},
\end{align}\end{subequations}
where $\phi=d/L$.  For $\phi=1$, that is for $d=L$ and the entire duct consists of the exponentially varying portion, the effective material properties reduce to the forms
\begin{subequations}\label{expTube}\begin{align}
	\kappa_\text{eff} &= \kappa_0\frac{mL}{e^{mL}-1}, \\
	\rho_\text{eff} &= \rho_0\frac{1-e^{-mL}}{mL}, \\
	\psi_\text{eff} &= \frac{\rho_0}{m}\frac{mL-\sinh(mL)}{e^{mL}-1}.
\end{align}\end{subequations}

\section{Conclusions}\label{conclusions}

This paper has developed and demonstrated a one-dimensional homogenization method based on transmission line theory.  Effective material properties, including the mass density, bulk modulus (or other one-dimensional measures of stiffness), and Willis coupling, may be readily evaluated in the long-wavelength limit.  The homogenization method has been formulated for both discrete systems and systems that vary smoothly in space.  The discrete homogenization method was demonstrated by considering the reflection of a plane acoustic pressure wave from a rigidly-backed bi-layer wall, and the reflection from an effective single-layer wall.  The true and effective reflection coefficients were shown to be equal at lowest order in frequency, and by including Willis coupling the effective reflection coefficient was shown to better approximate the true reflection coefficient at higher frequencies.  Finally, the continuous homogenization formulation was demonstrated by considering an exponentially growing horn embedded in an otherwise-uniform duct.

\appendix

\section{Product of Near-Identity Matrices}\label{appendixA}

Consider two matrices, $\tens A$ and $\tens B$, that are given by
\begin{subequations}\begin{align}
	\tens A = \tens I + \tens A_1 + \tens A_2, \\
	\tens B = \tens I + \tens B_1 + \tens B_2,
\end{align}\end{subequations}
where $\tens A_1$ and $\tens B_1$ are $O(\eps)$ and $\tens A_2$ and $\tens B_2$ are $O(\eps^2)$ for some $\eps\ll1$.  The product of these two matrices may then be written as
\begin{equation}
	\tens A\tens B = \tens I + \left[ \tens A_1 + \tens B_1 \right] + \left[ \tens A_2 + \tens B_2 + \tens A_1\tens B_1 \right] + O(\eps^3).
\end{equation}
Multiplying a third matrix with similar form $\tens C=\tens I +\tens C_1+\tens C_2$ yields
\begin{equation}
	\tens A\tens B\tens C = \tens I + \left[ \tens A_1 + \tens B_1 + \tens C_1 \right] + \left[ \tens A_2 + \tens B_2 + \tens C_2 + \tens A_1\tens B_1 + \tens A_1\tens C_1 + \tens B_1\tens C_1 \right] + O(\eps^3).
\end{equation}
Inductively, we conclude that for the product
\begin{equation}
	\Pi = \prod_{n=1}^N\tens A^{(n)} = \prod_{n=1}^N\left(\tens I + \tens A^{(n)}_1 + \tens A^{(n)}_2 \right)
\end{equation}
where $\tens A^{(n)}_1=O(\eps)$ and $\tens A^{(n)}_2=O(\eps^2)$ we may write
\begin{equation}
	\Pi = \tens I + \Pi_1 + \Pi_2 + O(\eps^3)
\end{equation}
where
\begin{subequations}\begin{align}
	\Pi_1 &= \sum_{n=1}^N \tens A^{(n)}_1 = O(\eps), \\
	\Pi_2 &= \sum_{n=1}^N \tens A^{(n)}_2 + \sum_{n=1}^{N-1}\sum_{m=n+1}^N\tens A^{(n)}_1\tens A^{(m)}_1 = O(\eps^2).
\end{align}\end{subequations}

\begin{acknowledgments}
	This research was supported by the U.\ S.\ Army Engineer Research and Development Center (ERDC), Environmental Quality and Installations business area.  Permission to publish was granted by Director, Cold Regions Research and Engineering Laboratory.
\end{acknowledgments}


\begin{thebibliography}{10}
\def\enquote#1,{``#1,''}
\expandafter\ifx\csname url\endcsname\relax
  \def\url#1{\texttt{#1}}\fi
\expandafter\ifx\csname urlprefix\endcsname\relax\def\urlprefix{URL }\fi
\providecommand{\bibinfo}[2]{#2}
\def\plainquote#1{``#1''}
\providecommand{\noopsort}[1]{}
\providecommand{\switchargs}[2]{#2#1}
\providecommand{\dourl}[1]{\href{http://#1}{\nolinkurl{#1}}}
\providecommand{\dodoi}[1]{doi: \href{http://dx.doi.org/#1}{\nolinkurl{#1}}}
  \def\eatspace #1{#1}

\bibitem{white1955}
\bibinfo{author}{J.~E. White} and \bibinfo{author}{F.~A. Angona},
  \enquote{\bibinfo{title}{Elastic {{Wave Velocities}} in {{Laminated
  Media}}}}, \bibinfo{journal}{The Journal of the Acoustical Society of
  America} \textbf{27}(2), \bibinfo{pages}{310--317} (\bibinfo{year}{1955})
  \dodoi{10.1121/1.1907520}.

\bibitem{rytov1956}
\bibinfo{author}{S.~M. Rytov}, \enquote{\bibinfo{title}{Acoustical properties
  of a thinly laminated medium}}, \bibinfo{journal}{Sov. Phys. Acoust.}
  \textbf{2}(1), \bibinfo{pages}{68--80} (\bibinfo{year}{1956}).

\bibitem{brekhovskikh2012}
\bibinfo{author}{L.M.~Brekhovskikh} and \bibinfo{author}{O.A.~Godin}, \emph{\bibinfo{title}{Acoustics of Layered Media}}, Vol.~\bibinfo{volume}{1}  (\bibinfo{publisher}{{Springer-Verlag}},
  \bibinfo{year}{1998}).



\bibitem{smith2011}
\bibinfo{author}{J.~D. Smith}, \enquote{\bibinfo{title}{Application of the
  method of asymptotic homogenization to an acoustic metafluid}},
  \bibinfo{journal}{Proc. R. Soc. A} \textbf{467}(2135),
  \bibinfo{pages}{3318--3331} (\bibinfo{year}{2011})
  \dodoi{10.1098/rspa.2011.0231}.

\bibitem{liu2009}
\bibinfo{author}{L.~Liu} and \bibinfo{author}{K.~Bhattacharya},
  \enquote{\bibinfo{title}{Wave propagation in a sandwich structure}},
  \bibinfo{journal}{International Journal of Solids and Structures}
  \textbf{46}(17), \bibinfo{pages}{3290--3300} (\bibinfo{year}{2009})
  \dodoi{10.1016/j.ijsolstr.2009.04.023}.

\bibitem{willis2009}
\bibinfo{author}{J.~R. Willis}, \enquote{\bibinfo{title}{Exact effective
  relations for dynamics of a laminated body}}, \bibinfo{journal}{Mechanics of
  Materials} \textbf{41}(4), \bibinfo{pages}{385--393} (\bibinfo{year}{2009})
  \dodoi{10.1016/j.mechmat.2009.01.010}.

\bibitem{brekhovskikh2012}
\bibinfo{author}{L.~Brekhovskikh}, \emph{\bibinfo{title}{Waves in Layered
  Media}}, Vol.~\bibinfo{volume}{16}  (\bibinfo{publisher}{{Elsevier}},
  \bibinfo{year}{2012}).

\bibitem{bobrovnitskii2014}
\bibinfo{author}{Y.~I. Bobrovnitskii}, \enquote{\bibinfo{title}{Effective
  parameters and energy of acoustic metamaterials and media}},
  \bibinfo{journal}{Acoust. Phys.} \textbf{60}(2), \bibinfo{pages}{134--141}
  (\bibinfo{year}{2014}) \dodoi{10.1134/S1063771014020018}.

\bibitem{jimenez2016}
\bibinfo{author}{N.~Jim\'enez}, \bibinfo{author}{V.~{Romero-Garc\'ia}},
  \bibinfo{author}{A.~Cebrecos}, \bibinfo{author}{R.~Pic\'o},
  \bibinfo{author}{V.~J. {S\'anchez-Morcillo}}, and \bibinfo{author}{L.~M.
  {Garcia-Raffi}}, \enquote{\bibinfo{title}{Broadband quasi perfect absorption
  using chirped multi-layer porous materials}}, \bibinfo{journal}{AIP Advances}
  \textbf{6}(12), \bibinfo{pages}{121605} (\bibinfo{year}{2016})
  \dodoi{10.1063/1.4971274}.

\bibitem{kutsenko2015}
\bibinfo{author}{A.~A. Kutsenko}, \bibinfo{author}{A.~L. Shuvalov},
  \bibinfo{author}{O.~Poncelet}, and \bibinfo{author}{A.~N. Darinskii},
  \enquote{\bibinfo{title}{Tunable effective constants of the one-dimensional
  piezoelectric phononic crystal with internal connected electrodes}},
  \bibinfo{journal}{The Journal of the Acoustical Society of America}
  \textbf{137}(2), \bibinfo{pages}{606--616} (\bibinfo{year}{2015})
  \dodoi{10.1121/1.4906162}.

\bibitem{sieck2015}
\bibinfo{author}{C.~F. Sieck}, \bibinfo{author}{A.~Al\`u}, and
  \bibinfo{author}{M.~R. Haberman}, \enquote{\bibinfo{title}{Dynamic
  {{Homogenization}} of {{Acoustic Metamaterials}} with {{Coupled Field
  Response}}}}, \bibinfo{journal}{Physics Procedia} \textbf{70},
  \bibinfo{pages}{275--278} (\bibinfo{year}{2015})
  \dodoi{10.1016/j.phpro.2015.08.153}.

\bibitem{baird1999}
\bibinfo{author}{A.~M. Baird}, \bibinfo{author}{F.~H. Kerr}, and
  \bibinfo{author}{D.~J. Townend}, \enquote{\bibinfo{title}{Wave propagation in
  a viscoelastic medium containing fluid-filled microspheres}},
  \bibinfo{journal}{The Journal of the Acoustical Society of America}
  \textbf{105}(3), \bibinfo{pages}{1527--1538} (\bibinfo{year}{1999})
  \dodoi{10.1121/1.426692}.

\bibitem{muhlestein2016}
\bibinfo{author}{M.~B. Muhlestein} and \bibinfo{author}{M.~R. Haberman},
  \enquote{\bibinfo{title}{A micromechanical approach for homogenization of
  elastic metamaterials with dynamic microstructure}}, \bibinfo{journal}{Proc.
  R. Soc. A} \textbf{472}(2192), \bibinfo{pages}{20160438}
  (\bibinfo{year}{2016}) \dodoi{10.1098/rspa.2016.0438}.

\bibitem{zhu2015}
\bibinfo{author}{R.~Zhu}, \bibinfo{author}{X.~N. Liu}, \bibinfo{author}{G.~K.
  Hu}, \bibinfo{author}{F.~G. Yuan}, and \bibinfo{author}{G.~L. Huang},
  \enquote{\bibinfo{title}{Microstructural designs of plate-type elastic
  metamaterial and their potential applications: A review}},
  \bibinfo{journal}{International Journal of Smart and Nano Materials}
  \textbf{6}(1), \bibinfo{pages}{14--40} (\bibinfo{year}{2015})
  \dodoi{10.1080/19475411.2015.1025249}.

\bibitem{muhlestein2017a}
\bibinfo{author}{M.~B. Muhlestein}, \bibinfo{author}{C.~F. Sieck},
  \bibinfo{author}{P.~S. Wilson}, and \bibinfo{author}{M.~R. Haberman},
  \enquote{\bibinfo{title}{Experimental evidence of {{Willis}} coupling in a
  one-dimensional effective material element}}, \bibinfo{journal}{Nat Commun}
  \textbf{8} (\bibinfo{year}{2017}) \dodoi{10.1038/ncomms15625}.

\bibitem{muhlestein2016a}
\bibinfo{author}{M.~B. Muhlestein}, \bibinfo{author}{C.~F. Sieck},
  \bibinfo{author}{A.~Al\`u}, and \bibinfo{author}{M.~R. Haberman},
  \enquote{\bibinfo{title}{Reciprocity, passivity and causality in {{Willis}}
  materials}}, \bibinfo{journal}{Proceedings of the Royal Society of London A}
  \textbf{472}(2194), \bibinfo{pages}{20160604} (\bibinfo{year}{2016})
  \dodoi{10.1098/rspa.2016.0604}.

\end{thebibliography}

\end{document}